# Harnessing the Digital Revolution: A Comprehensive Review of mHealth Applications for Remote Monitoring in Transforming Healthcare Delivery

Avnish Singh Jat[1(✉)] [0000-0001-7473-3010] and Tor-Morten Grønli [1] [0000-0002-2026-4551]

[1] School of Economics, Innovation, and Technology
Kristiania University College, Oslo, Norway
`avnishsingh.jat@kristiania.no`

**Abstract.** The utilization of mHealth applications for remote monitoring has the potential to revolutionize healthcare delivery by enhancing patient outcomes, increasing access to healthcare services, and reducing healthcare costs. This literature review aims to provide a comprehensive overview of the current state of knowledge on mHealth applications for remote monitoring, including their types, benefits, challenges, and limitations, as well as future directions and research gaps. A systematic search of databases such as PubMed, MEDLINE, EMBASE, CINAHL, and Google Scholar was conducted to identify relevant articles published within the last 5 years. Thematic analysis was used to synthesize the findings.

The review highlights various types of mHealth applications used for remote monitoring, such as telemedicine platforms, mobile apps for chronic disease management, and wearable devices. The benefits of these applications include improved patient outcomes, increased access to healthcare, reduced healthcare costs, and addressing healthcare disparities. However, challenges and limitations, such as privacy and security concerns, lack of technical infrastructure, regulatory issues, data accuracy, user adherence, and the digital divide, need to be addressed to ensure successful adoption and utilization of mHealth applications.

Further research is required in areas such as the long-term effects of mHealth applications on patient outcomes, integration of mHealth data with electronic health records, and the development of artificial intelligence-driven mHealth applications. By harnessing the potential of mHealth applications and addressing the existing challenges, healthcare delivery can be transformed towards a more accessible, cost-effective, and patient-centered model.

## 1   Introduction

The rapid advancement of technology has transformed various aspects of our lives, and healthcare is no exception. The intersection of healthcare and technology has led to the development of innovative solutions that can improve healthcare delivery, enhance patient outcomes, and reduce costs. One such innovation is the use of mobile health (mHealth) applications for remote monitoring. This literature review aims to explore the current state of knowledge on the utilization of mHealth applications for remote monitoring and to identify research gaps and future directions in this field.

mHealth applications refer to the use of mobile and wireless devices, such as smartphones, tablets, and wearable sensors, to support medical and public health practices. These applications enable the collection, transmission, analysis, and presentation of health-related data, facilitating better healthcare management and decision-making.[1] Remote monitoring, on the other hand, is a healthcare delivery method that allows healthcare providers to continuously monitor patients' health status and provide timely interventions without the need for in-person visits. Remote monitoring is particularly relevant for managing chronic diseases, such as diabetes, hypertension, and heart failure, where continuous monitoring and timely interventions are crucial for improving patient outcomes.[2] The significance of mHealth applications for remote monitoring in healthcare delivery cannot be overstated. As the global population ages and the prevalence of chronic diseases increases, there is a growing demand for cost-effective, accessible, and patient-centered healthcare solutions.[3] mHealth applications have the potential to address these needs by providing real-time monitoring, personalized feedback, and enhanced communication between patients and healthcare providers.[4] Furthermore, these applications can help bridge the gap in healthcare access, particularly in rural and low-resource settings, where traditional healthcare delivery methods may be limited.[5]

The purpose of this literature review is to synthesize existing research on the utilization of mHealth applications for remote monitoring and to identify research gaps and future directions in this field. Specifically, the literature review seeks to answer the following research questions:

a) *What types of mHealth applications are currently being used for remote monitoring, and what are their key features and functionalities?*
b) *What are the benefits of using mHealth applications for remote monitoring in terms of patient outcomes, healthcare access, and cost reduction?*
c) *What are the challenges and limitations associated with the adoption and utilization of mHealth applications for remote monitoring?*
d) *What are the potential solutions to overcome these challenges and limitations, and what are the emerging trends and technologies that could shape the future of mHealth applications for remote monitoring?*

By examining the existing literature on mHealth applications for remote monitoring, this review aims to provide a comprehensive understanding of the current state of knowledge in this field, inform future research, and contribute to the ongoing discourse on the role of technology in transforming healthcare delivery.

## 2    Background

The development of mHealth applications and the growth of remote monitoring in healthcare have evolved over the past few decades, driven by the convergence of various technological advancements and the growing need for more accessible and cost-effective healthcare solutions. This section provides an overview of the evolution of mHealth applications and the factors contributing to their increased adoption in remote monitoring.

The origins of mHealth can be traced back to the 1990s, when mobile devices were first used for basic healthcare purposes, such as sending short message service (SMS) reminders for medication adherence or appointment scheduling.[6] With the advent of smartphones in the late 2000s, mHealth applications started to gain traction as these devices offered enhanced processing capabilities, larger screens, and an expanding ecosystem of applications.[7] Around the same time, the proliferation of wearable devices and sensors capable of collecting and transmitting health-related data further fueled the development of mHealth applications.[8]

The growth of remote monitoring in healthcare can be attributed to several factors, including the rising prevalence of chronic diseases, an aging global population, and the increasing burden on healthcare systems. Remote monitoring has emerged as a viable solution to address these challenges, as it allows healthcare providers to monitor patients' health status continuously and intervene in a timely manner, thus improving outcomes and reducing hospitalizations.[9]

There are several key factors driving the increased adoption of mHealth applications for remote monitoring in healthcare:

Technological advancements: The rapid evolution of mobile and wireless technologies, such as smartphones, wearable devices, and sensors, has enabled the development of sophisticated mHealth applications that can collect, analyze, and transmit health-related data in real-time. Additionally, advancements in data analytics, artificial intelligence, and cloud computing have further enhanced the capabilities of mHealth applications in remote monitoring.[10]

Growing smartphone penetration: The widespread adoption of smartphones has made it easier for both patients and healthcare providers to access mHealth applications. According to the Pew Research Center, smartphone ownership has grown significantly over the past decade, with more than 5 billion people worldwide now using smartphones. This increased accessibility has facilitated the adoption of mHealth applications for remote monitoring across various demographic groups and geographical regions.[11]

Need for cost-effective healthcare solutions: The rising costs of healthcare have become a significant concern for governments, healthcare providers, and patients alike. mHealth applications for remote monitoring offer a cost-effective solution by reducing the need for in-person visits, minimizing hospital readmissions, and enabling more efficient use of healthcare resources. Studies have shown that mHealth applications can lead to significant cost savings, particularly in the management of chronic diseases.[12][13]

Patient empowerment and demand for personalized care: There has been a growing trend towards patient empowerment and a demand for personalized healthcare solutions. mHealth applications for remote monitoring empower patients to actively participate in their healthcare management and enable healthcare providers to deliver personalized interventions based on real-time data. This shift towards patient-centered care has contributed to the increased adoption of mHealth applications for remote monitoring.[14]

The development of mHealth applications and the growth of remote monitoring in healthcare have been driven by a combination of technological advancements, increasing smartphone penetration, and the need for cost-effective and personalized healthcare solutions. As these trends continue to evolve, it is expected that mHealth applications for remote monitoring will play an increasingly critical role in healthcare delivery.

# 3   Methodology

This section provides an overview of the approach taken to collect, analyze, and synthesize the relevant research on the utilization of mHealth applications for remote monitoring. The methodology section will discuss the search strategy, inclusion and exclusion criteria, data extraction, and analysis process.

### 3.1. Search Strategy
To identify relevant articles and studies on mHealth applications for remote monitoring, a comprehensive search of databases such as PubMed, MEDLINE, EMBASE, CINAHL, and Google Scholar was conducted. The search terms used included combinations of keywords such as "mHealth," "mobile health," "remote monitoring," "telemedicine," "digital health," and "healthcare applications." Additionally, reference lists of the included articles were reviewed to identify any other relevant publications.

### 3.2.   Inclusion and Exclusion Criteria
To ensure the relevance and quality of the literature included in the review, specific inclusion and exclusion criteria were established. Studies were included if they:
   a) Were published in English.
   b) Focused on mHealth applications used for remote monitoring in healthcare settings.
   c) Presented primary research, such as clinical trials, observational studies, or qualitative research.

Were published within the last 5 years to ensure the relevance of the technology and findings.

Exclusion criteria were applied to remove studies that:
   a) Did not focus on mHealth applications or remote monitoring.
   b) Were reviews, editorials, or opinion pieces without primary data.
   c) Focused on telehealth interventions that did not involve mobile or digital technology.

### 3.3. Data Extraction
After the identification of relevant articles and studies, data were extracted using a standardized data extraction form. The extracted information included the study's objectives, design, population, mHealth application type, intervention details, outcome measures, and main findings.

### 3.4. Data Analysis
The extracted data were analyzed and synthesized using a thematic approach. This involved identifying common themes, patterns, and trends across the included studies.[15] The analysis was organized into sections that addressed the different aspects of mHealth applications for remote monitoring, such as types of applications, benefits, challenges and limitations, future directions, and research gaps.

By following a rigorous methodology, the literature review aims to provide a comprehensive and up-to-date overview of the current state of knowledge on the utilization of mHealth applications for remote monitoring in healthcare. This approach ensures that the review's findings are based on a thorough examination of the available evidence, which can serve as a foundation for further research and inform policy and practice decisions in this rapidly evolving field.

## 4 mHealth Applications for Remote Monitoring

mHealth applications for remote monitoring encompass a wide range of tools and platforms designed to facilitate healthcare delivery and improve patient outcomes. These applications as per our review can be broadly categorized into three main types: telemedicine platforms, mobile apps for chronic disease management, and wearable devices. This section will discuss each of these categories, their key features and functionalities, and examples of successful mHealth applications used in various medical fields.

### 4.1 Telemedicine platforms
Telemedicine platforms allow healthcare providers to conduct virtual consultations with patients, enabling remote diagnosis, treatment planning, and follow-up care. These platforms often include features such as video conferencing, secure messaging, electronic health record (EHR) integration, and appointment scheduling.[16] Telemedicine platforms have become particularly useful during the COVID-19 pandemic, as they have enabled healthcare providers to continue delivering care while minimizing the risk of infection.[17]

Practo is a leading telemedicine platform in India that connects patients with healthcare providers through video consultations, allowing for remote diagnosis and treatment of various medical conditions, including dermatological issues, mental health concerns, and chronic disease management.[18]

**4.2 Mobile apps for chronic disease management**
Mobile apps for chronic disease management are designed to help patients and healthcare providers monitor and manage chronic conditions, such as diabetes, hypertension, and heart failure. These apps typically include features such as medication reminders, symptom tracking, personalized feedback, and educational resources. Some apps also integrate with wearable devices or external sensors to collect real-time health data, which can be shared with healthcare providers for analysis and intervention.[19][20]
MySugr is a mobile app designed for diabetes management, allowing patients to track their blood glucose levels, insulin doses, and carbohydrate intake. The app provides personalized feedback and insights to help users maintain optimal blood sugar control and integrates with various glucose meters and insulin pumps for seamless data collection.[21]

**4.3 Wearable devices**
Wearable devices, such as fitness trackers and smartwatches, are increasingly being used for remote monitoring purposes, as they can collect and transmit various health-related data, including heart rate, physical activity, sleep patterns, and more. Some wearable devices also incorporate specialized sensors to monitor specific health parameters, such as blood glucose levels or blood pressure, making them particularly useful for managing chronic diseases.[10][22]
Apple Watch Series 7 features an FDA-cleared electrocardiogram (ECG) app that can detect atrial fibrillation, a common form of irregular heartbeat. Users can share their ECG data with their healthcare providers, allowing for remote monitoring and timely interventions when necessary.[23]

**4.4 Features of mHealth Applications**
The key features and functionalities of mHealth applications for remote monitoring include:
Data collection: mHealth applications collect various types of health-related data, such as vital signs, symptoms, medication adherence, and lifestyle factors, to provide a comprehensive view of a patient's health status.[24]
Patient-provider communication: These applications often facilitate secure communication between patients and healthcare providers, enabling remote consultations, real-time feedback, and timely interventions.[25]
Data analysis: mHealth applications typically incorporate data analytics and artificial intelligence capabilities to analyze health data, identify trends and patterns, and generate personalized insights and recommendations.[24]
mHealth applications for remote monitoring have been successfully used in various medical fields, including:
Cardiology: AliveCor's KardiaMobile is a portable ECG device that connects to a smartphone app, enabling patients to monitor their heart rhythm and share the data with their healthcare providers for remote monitoring and intervention.[26]
Diabetes management: Dexcom G6 is a continuous glucose monitoring (CGM) system that tracks blood glucose levels in real-time, sends alerts when levels are too high or low, and shares the data with healthcare providers for remote monitoring and personalized feedback.[27]

Mental health: Headspace is a meditation and mindfulness app that offers guided sessions, tools, and resources to help users manage stress, anxiety, and depression. While Headspace does not directly provide remote monitoring by healthcare providers, the app allows users to track their progress and share their mental health data with therapists or other healthcare professionals, facilitating remote support and personalized interventions. Headspace has been shown to be effective in reducing stress, improving focus, and promoting overall mental well-being.[28]

mHealth applications for remote monitoring encompass a diverse range of tools and platforms, such as telemedicine platforms, mobile apps for chronic disease management, and wearable devices. These applications offer various features and functionalities, including data collection, patient-provider communication, and data analysis, to facilitate remote healthcare delivery and improve patient outcomes. Examples of successful mHealth applications in various medical fields demonstrate the potential of these technologies in transforming healthcare delivery and addressing the growing demand for cost-effective, accessible, and patient-centered care.

## 5 Benefits of mHealth Applications for Remote Monitoring

The utilization of mHealth applications for remote monitoring offers several advantages that can transform healthcare delivery and improve patient outcomes. This section will discuss the key benefits of mHealth applications for remote monitoring, review the evidence supporting their effectiveness, and explore their potential to address healthcare disparities in rural and low-resource settings.

Improved patient outcomes

mHealth applications for remote monitoring enable continuous monitoring of patients' health status, allowing healthcare providers to intervene in a timely manner and prevent complications. This is particularly beneficial for patients with chronic conditions, where early detection and intervention can significantly improve outcomes.[29] Several studies and trials have demonstrated the effectiveness of mHealth applications in improving patient outcomes. A study published in JAMA found that patients using a mobile app for heart failure management experienced fewer hospitalizations and improved quality of life compared to those receiving standard care.[30]

Increased access to healthcare

mHealth applications can bridge geographical barriers and provide access to healthcare services for individuals in remote or underserved areas. Telemedicine platforms, for instance, enable virtual consultations with healthcare providers, reducing the need for patients to travel long distances for in-person visits.[31] Moreover, mHealth applications can facilitate access to specialized care for patients with rare or complex conditions, who may not have access to expert care in their local communities. A systematic review published in the Journal of Medical Internet Research found that telemedicine interventions can improve access to healthcare and patient satisfaction, particularly in rural and remote areas.[32]

Reduced healthcare costs
By enabling remote monitoring and reducing the need for in-person visits, mHealth applications can lead to significant cost savings for both patients and healthcare providers.[33] A study published in Journal of Medical Internet Research: Cardiovascular Quality and Outcomes found that a telemonitoring program for heart failure patients resulted in a 30% reduction in hospitalization costs.[34] Furthermore, mHealth applications can streamline healthcare processes, such as appointment scheduling and medication management, leading to increased efficiency and reduced administrative costs.[33]

Addressing healthcare disparities
mHealth applications have the potential to address healthcare disparities by providing access to healthcare services for traditionally underserved populations, such as those in rural or low-resource settings. By leveraging mobile and wireless technologies, mHealth applications can overcome infrastructure limitations, such as the lack of healthcare facilities and healthcare providers and offer cost-effective healthcare solutions for these populations.[35] A study published in the International Journal of Medical Informatics found that a mobile health intervention targeting rural populations with chronic diseases led to improved medication adherence and disease-specific knowledge.[36]

mHealth applications for remote monitoring offer several advantages, including improved patient outcomes, increased access to healthcare, reduced healthcare costs, and the potential to address healthcare disparities. The evidence from studies and trials supports the effectiveness of mHealth applications in improving patient outcomes and healthcare delivery, particularly for individuals in rural and low-resource settings. As technology continues to advance and mHealth applications become more sophisticated, their role in transforming healthcare delivery and addressing healthcare disparities is expected to grow.

## 6      Challenges and Limitations

While mHealth applications for remote monitoring offer numerous benefits, they also face several challenges and limitations that can hinder their adoption and utilization. This section will identify the key barriers, discuss the limitations of mHealth applications, and explore potential solutions to overcome these challenges.

Privacy and security concerns
One of the primary challenges associated with mHealth applications is ensuring the privacy and security of sensitive health information. Patients and healthcare providers may be hesitant to adopt mHealth applications due to concerns about unauthorized access to personal health data, data breaches, or misuse of information.[37] To address these concerns, developers of mHealth applications must adhere to strict data protection standards, implement robust encryption methods, and comply with relevant regulations, such as the Health Insurance Portability and Accountability Act (HIPAA) in the United States.[38] Additionally, one research reviewed mentioned the integration of technologies like blockchain for security can help enhance data protection and increase confidence in the use of mHealth applications for remote monitoring.[39]

Lack of technical infrastructure
The successful implementation of mHealth applications for remote monitoring requires a robust technical infrastructure, including reliable internet connectivity, compatible devices, and interoperable systems. In some rural or low-resource settings, the lack of adequate infrastructure can hinder the adoption of mHealth applications. To overcome this challenge, governments and organizations should invest in improving connectivity and infrastructure in underserved areas, while developers should design mHealth applications that can function with limited resources or in offline mode.[40][41]

Regulatory issues
Navigating the complex regulatory landscape for mHealth applications can be challenging for developers and healthcare providers alike. Different countries and regions have varying regulations regarding the approval, use, and reimbursement of mHealth applications, which can create confusion and slow down the adoption process. To address this issue, stakeholders should advocate for harmonized regulatory frameworks and clear guidelines for the development and deployment of mHealth applications.[42][43]

Data accuracy
Ensuring the accuracy and reliability of health data collected through mHealth applications is critical for effective remote monitoring. Some wearable devices or sensors may not be as accurate as traditional medical devices, which can compromise the quality of care provided. To overcome this limitation, developers should prioritize the validation and calibration of mHealth applications and devices, while healthcare providers should be cautious when interpreting data from these sources.[44][45]

User adherence
The effectiveness of mHealth applications for remote monitoring largely depends on user adherence, which can be influenced by factors such as usability, user engagement, and perceived value. To enhance user adherence, developers should design mHealth applications that are user-friendly, customizable, and incorporate gamification or motivational strategies to encourage sustained use.[46]

Digital divide
The digital divide, or the gap between those with access to digital technologies and those without, can limit the reach of mHealth applications for remote monitoring. Individuals from low-income, elderly, or rural populations may face barriers in accessing or using mHealth applications due to factors such as affordability, digital literacy, or lack of infrastructure. To bridge the digital divide, stakeholders should promote digital literacy initiatives, subsidize the cost of devices or internet access, and develop low-cost mHealth solutions tailored to the needs of these populations.[47]

Addressing the challenges and limitations of mHealth applications for remote monitoring is essential to ensure their successful adoption and utilization. By developing standardized protocols, promoting digital literacy, ensuring data privacy, and investing in infrastructure improvements, stakeholders can help overcome these barriers and unlock the full potential of mHealth applications in transforming healthcare delivery and improving patient outcomes.

# 7 Future Directions and Research Gaps

As mHealth applications for remote monitoring continue to evolve and gain traction, there are several areas where further research is needed to better understand their impact and potential. This section will highlight key research gaps and discuss emerging trends and technologies that could shape the future of mHealth applications for remote monitoring.

Integration of mHealth data with electronic health records
The seamless integration of mHealth data with electronic health records (EHRs) is crucial for optimizing healthcare delivery and facilitating data-driven decision-making. Research is needed to develop interoperable systems, data standards, and protocols that enable efficient and secure sharing of mHealth data across different healthcare providers and platforms.

Artificial intelligence-driven mHealth applications
The integration of artificial intelligence (AI) and machine learning algorithms into mHealth applications holds significant promise for enhancing remote monitoring capabilities, personalizing care, and predicting patient outcomes. Further research is needed to develop and validate AI-driven mHealth applications, as well as to understand the ethical, legal, and social implications of their use in healthcare.

Internet of Things (IoT) and remote monitoring
The Internet of Things (IoT) refers to the interconnected network of devices and sensors that can collect, share, and analyze data. IoT has the potential to transform mHealth applications for remote monitoring by enabling the seamless integration of wearable devices, home monitoring systems, and environmental sensors. Research is needed to develop IoT-based mHealth applications, evaluate their effectiveness, and address the associated privacy and security challenges.

Advanced data analytics

As mHealth applications generate vast amounts of health data, there is a growing need for advanced data analytics techniques to process, analyze, and interpret this information. Research is needed to develop novel analytics algorithms and tools that can identify trends, patterns, and correlations in mHealth data, enabling healthcare providers to make more informed decisions and deliver personalized care.

## 8 Conclusion

In conclusion, mHealth applications for remote monitoring have the potential to transform healthcare delivery by offering several benefits, including improved patient outcomes, increased access to healthcare services, reduced healthcare costs, and addressing healthcare disparities. Despite the challenges and limitations, such as privacy and security concerns, lack of technical infrastructure, regulatory issues, data accuracy, user adherence, and the digital divide, potential solutions like standardized protocols, digital literacy promotion, and ensuring data privacy can help overcome these barriers.

As the field continues to grow, further research is required to address gaps in areas like long-term effects on patient outcomes, integration of mHealth data with electronic health records, and the development of artificial intelligence-driven mHealth applications. Emerging trends and technologies, such as 5G, the Internet of Things, and advanced data analytics, will shape the future of mHealth applications for remote monitoring, enabling more sophisticated and personalized care.

Collaboration among stakeholders, ongoing research, and investment in infrastructure improvements will be crucial to ensure that mHealth applications remain responsive to the changing needs of patients and healthcare providers. By harnessing the potential of mHealth applications for remote monitoring, we can work towards a future where healthcare is more accessible, cost-effective, and patient-centered.

12